\documentclass[aps,prb,twocolumn,superscriptaddress,showpacs,floatfix]{revtex4}

\usepackage{epsfig}

\begin{document}

\title{Self field of ac current reveals
voltage--current law in type-II superconductors}

\author{E.~H.~Brandt}
\affiliation{Max-Planck-Institut f\"ur Metallforschung,
   D-70506 Stuttgart, Germany}

\author{G.~P.~Mikitik}
\affiliation{Max-Planck-Institut f\"ur Metallforschung,
   D-70506 Stuttgart, Germany}
\affiliation{B.~Verkin Institute for Low Temperature Physics
   \& Engineering, Ukrainian Academy of Sciences,
   Kharkov 61103, Ukraine}

\author{E.~Zeldov}
\affiliation{Department of Condensed Matter Physics,
Weizmann Institute of Science, Rehovot 76100, Israel}

\date{\today}

\begin{abstract}
The distribution of ac magnetic fields in a thin superconducting
strip is calculated when an ac current is applied to the sample.
The edge barrier of the strip to flux penetration and exit is
taken into account. The obtained formulas provide a basis to
extract voltage-current characteristics in the bulk and at the
edges of superconductors from the measured ac magnetic fields.
Based on these results, we explain the spatial and
temperature dependences of the profiles of the ac magnetic field
measured in a Bi$_2$Sr$_2$CaCu$_2$O$_8$ strip [D.T.\ Fuchs {et
al}., Nature {\bf 391}, 373 (1998)].
\end{abstract}


\maketitle

\section{Introduction}   

Recently a novel experimental approach for mapping the
distribution of the transport current across a flat
superconducting sample was developed. \cite{1,2} In this technique
an array of microscopic Hall sensors is attached directly to the
sample and the perpendicular component of the magnetic self-field
generated by an ac transport current is measured at various
locations across the sample. Inverting the Biot-Savart law, the
distribution of the transport current across the sample can be
obtained from the measured self field. It was found that in
contrast to the common assumptions the transport-current
distribution is highly nonuniform and varies significantly as a
function of temperature $T$, the applied magnetic field $H_a$, and
the phase of the vortex matter, Fig.~1. Surprisingly, over a wide
range of the $H$-$T$ phase diagram the current flows predominantly
{\em at the edges} of the sample due to significant surface
barriers both in high-$T_c$ and clean low-$T_c$ superconductors.
\cite{1,2,3} In this situation the velocity of the vortex lattice
is determined by the rate of the vortex activation over the
surface barrier rather than by the bulk vortex pinning. At lower
temperatures bulk vortex dynamics takes over, resulting in
redistribution of the transport current; see Fig.~1. So far,
however, the experimental data thus obtained were not used for
quantitative study of vortex dynamics due to the absence of a
detailed theoretical description.

In this paper we develop such a description. It provides the basis
for the extraction of various voltage-current characteristics in
the bulk and at the edges of superconductors. In Sec.~II we give
general formulas for the ac magnetic fields and currents, while in
Secs.~III-V three simple models are considered which shed light on
the data presented in Fig.~1. Combinations of these models (or of
more complicated models developed on the basis of the general
equations of Sec.~II) enable one to analyze various experimental
situations in different superconductors. The obtained results show
that the self-field method can be a useful tool for determination
of the voltage--current characteristics of superconductors. This
development supplements the usual transport measurements and thus
opens novel possibilities for a quantitative investigation of
vortex dynamics.

 \begin{figure}  
 \includegraphics[scale=.47]{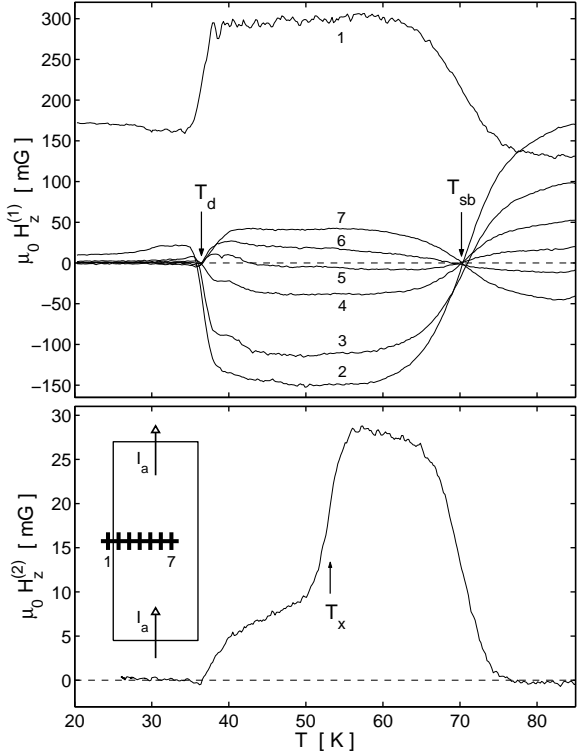}
\caption{\label{fig1} Top: The temperature dependence of the
in-phase part of the first harmonic of the ac self field measured
by an array of 7 Hall probes at the surface of a
Bi$_2$Sr$_2$CaCu$_2$O$_8$ strip carrying an ac current of 4 mA
amplitude in a constant applied dc field of 1000 G. Bottom: The
temperature dependence of the second harmonic of the ac magnetic
field measured by one of the sensors at 10 mA ac current at 1000
G. Shown are also the characteristic temperatures $T_d$, $T_{sb}$,
$T_x$. All the data from Ref.~\onlinecite{2}. The inset shows the
geometry of the experiment and the positions of the sensors.
 } \end{figure}   

\section{General formulas for AC magnetic field and current} 

\subsection{Edge barrier}   

Consider an infinitely long thin superconducting strip of width
$2w$ and thickness $d \ll w$, filling the space $|x|\le w$, $|y|
<\infty $, $|z|\le d/2$. Let the strip be in a constant and
uniform external magnetic field $H_a$ directed along the $z$ axis,
and let an ac current $I_a=I_{a0}\sin \omega t$ of frequency
$\omega$ and of magnitude $I_{a0}$ be applied to the sample. The
current--voltage dependence for the superconducting strip in the
bulk is implied to have the standard form $E=\rho(|j|) j$, with
resistivity $\rho$ being a nonlinear function of the current
density $j$: At small current densities, $|j|<j_c$, the
resistivity $\rho$ (and thus the electric field $E$) is
practically equal to zero; near the critical current density $j_c$
(i.e., at $|j|\approx j_c$) the resistivity sharply increases, and
it reaches a constant value $\rho_{\rm ff}$ independent of $j$ in
the flux-flow regime at $|j|\gg j_c$.

In the narrow regions of the strip near its right and left edges
$x=\pm w$ where the barrier for flux penetration into the strip or
exit from it occurs, we shall use the current--voltage laws:
$E=R_rI_{r}$, $E=R_lI_{l}$ where $I_{r}$ and $I_{l}$ are the
currents that flow in these edge regions while $R_r$ and $R_l$ are
the appropriate resistances per unit length in $y$. In other
words, we model the small edge regions of the strip as ``wires''
with the resistances $R_r$ and $R_l$. Since the edge barrier is
generally different for flux entrance into the sample and for flux
exit from it, \cite{bkv} these resistances are generally
different, too. When $I_a>0$, the vortices enter the strip at
$x=-w$ and exits from it at $x=w$, i.e., one has $R_l=R_{en}$ and
$R_r=R_{ex}$, while at $I_a<0$ we arrive at the opposite
relations: $R_l=R_{ex}$ and $R_r=R_{en}$. In general, $R_{ex}$ and
$R_{en}$ are nonlinear functions of the currents flowing in the
appropriate edge regions. \cite{bkv}

We may estimate the radius of the two edge wires from the
following arguments. For a strip in a parallel magnetic field when
$H_a$ is applied along the $x$ axis and is larger than the
penetration field,  the $z$-size of the region where the surface
currents flow is less than the London penetration depth $\lambda$.
\cite{Clem} It is generally believed that under condition $\lambda
\ll d$, which usually holds for single crystals of high-$T_c$
superconductors, this is also true in the considered case of a
perpendicular magnetic field ($H_a\parallel z$) if $H_a$ is higher
than the lower critical field $H_{c1}$. In other words, one might
expect that the width of the edge wires is less or of the order of
$\lambda$. However, simple considerations show that this is not
the case since this assumption leads to a contradiction at least
for not-too-large ratio $H_a/H_{c1}$. Indeed, consider the
situation when flux line pinning is negligible. If the edge
currents flew only at $w\ge |x| \ge w-\lambda$, they would
generate both $H_z$ and $H_x$ at least in the region $w\ge |x| \ge
w-d$. This means that the flux lines are curved at such $x$. But
without pinning and currents the curved flux lines cannot be in
equilibrium. Thus, we conclude that in a strip in a perpendicular
magnetic field the edge currents flow in the whole region $w-|x|
\le d$ and probably have a complicated distribution over $x$ and
$z$ there. With this in mind we shall assume below that the width
of the edge wires is of the order of $d$. Note that for $H_a$ of
the order of the lower critical field $H_{c1}$ this assumption
agrees with the known result for the width of the geometrical
barrier. \cite{Z}

\subsection{Simple case}  

We now consider the magnetic field $H_z(x,t)$ generated by the ac
current $I_a=I_{a0}\sin \omega t$. In particular, we calculate the
first and the second harmonics of the ac magnetic field $H_z$:
\begin{eqnarray}\label{1a}
 H_z^{(1)}(x)=\frac{2}{T}\int_0^T H_z (x,t)\sin{\omega t}dt, \\
 H_z^{(2)}(x)=\frac{2}{T}\int_0^T H_z(x,t)\cos {2\omega t}dt, \label{1b}
\end{eqnarray}
where $T=2\pi/\omega$ is the period of the oscillating ac current.
(When the inductance of the strip is negligible, it follows from
symmetry considerations that the first harmonic proportional to
$\cos\omega t$ and the second harmonic proportional to $\sin
2\omega t$ are equal to zero.)

The total current $I_a$ flowing in the strip at time $t$ splits
into the bulk current $I_b$ and the edge current $I_e$,
 \begin{equation}\label{2}
 I_a=I_b+I_e,
 \end{equation}
where $I_e$ consists of the left ($I_{l}$) and right ($I_{r}$)
parts,
 \begin{equation}\label{3}
 I_e=I_{l}+I_{r}.
 \end{equation}
If the frequency $\omega$ is not too large, the current
distribution over the cross-section of the sample is uniform, and
hence the sheet current in the strip $J(x)$ (i.e., the current
density integrated over the thickness $d$ of the strip) is
independent of $x$,
 \begin{equation}\label{4}
 J(x)={I_b \over 2w}\,.
 \end{equation}
In this case we immediately obtain
 \begin{equation}\label{5}
 I_e=I_a {R_b \over R_b+R_e}\,,\ \ \ I_b=I_a {R_e \over R_b+R_e}\,,
 \end{equation}
where $R_b=\rho(J/d)/2wd$ and
 \begin{equation}\label{6}
 R_e={R_lR_r\over R_l+R_r}={R_{en}R_{ex}\over R_{en}+R_{ex}}\,.
 \end{equation}
As to $I_{l}$ and $I_{r}$, one finds
 \begin{equation}\label{7}
 I_{l}=I_e {R_r \over R_l+R_r}\,,\ \ \
 I_{r}=I_e {R_l \over R_l+R_r}\,,
 \end{equation}
with
 \begin{equation}\label{8}
 R_l=R_{en}\,,\ \ \  R_r=R_{ex}\ \ \ \  {\rm for}\  I_a>0,
 \end{equation}
and
 \begin{equation}\label{9}
 R_l=R_{ex}\,,\ \ \  R_r=R_{en}\ \ \ \  {\rm for}\  I_a<0.
 \end{equation}

As was mentioned above, $R_b$, $R_{en}$ and $R_{ex}$ are generally
nonlinear functions of $J=I_b/2w$ and of the currents $I_{en}$ and
$I_{ex}$, i.e., $R_b=R_b(I_b)$, $R_{en}=R_{en}(I_{en})$ and
$R_{ex}=R_{ex}(I_{ex})$, where $I_{en}=I_{l}$, $I_{ex}=I_{r}$ for
$I_a>0$, and $I_{en}=I_{r}$, $I_{ex}=I_{l}$ for $I_a<0$.
Therefore, formulas (\ref{5}) - (\ref{7}) are in fact equations
for $I_b$, $I_{l}$ and $I_{r}$,
 \begin{eqnarray}\label{10a}
I_{l}\!=\!I_a {R_r(I_{r}) R_b(I_b) \over [R_l(I_{l})+R_r(I_{r})]
R_b(I_b) + R_r(I_{r})R_l(I_{l})},~\\
I_{r}\!=\!I_a {R_l(I_{l}) R_b(I_b) \over [R_l(I_{l})+R_r(I_{r})]
R_b(I_b) + R_r(I_{r})R_l(I_{l})},~ \label{10b} \\
I_b\!=\!I_a {R_r(I_{r}) R_l(I_{l}) \over
[R_l(I_{l})\!+\!R_r(I_{r})] R_b(I_b)\!
 +\!R_r(I_{r})R_l(I_{l})}.~ \label{10c}
 \end{eqnarray}
On determining $I_{l}$, $I_{r}$, and $I_b$ from this set of
equations, one can calculate $H_z(x,t)$ as sum of the fields of
the wires and of the spatially uniform sheet current $J$,
 \begin{eqnarray}
 H_z(x,t)={1 \over 2\pi }\left ({I_b\over 2w}
 \ln \Big |{w-x\over w+x}\Big |
 + {I_{r}\over w-x}-{I_{l}\over w+x}
   \right ). \label{11}
 \end{eqnarray}
This formula together with Eqs.~(\ref{10a})-(\ref{10c}) gives
$H_z$ as a function of $x$ and $I_a$, $H_z=H_z(x,I_a)$. Expressing
the time $t$ in Eqs.~(\ref{1a}), ({\ref{1b}) via $I_a$,
$\sin\omega t=I_a/I_{a0}=\tilde I_a$, we arrive at
\begin{eqnarray}\label{12a}
 H_z^{(1)}(x)=\frac{4}{\pi}\int_0^1 H_z (x,\tilde I_a){\tilde I_a
 d\tilde I_a \over (1- \tilde I_a^2)^{1/2}},\ \  \\
 H_z^{(2)}(x)\!=\!\frac{2}{\pi}\!\int_0^1\!\![H_z(x,\tilde I_a\!)\!+
 \!H_z(x,-\tilde I_a\!)] {(1-2\tilde I_a^2)
 d\tilde I_a \over (1- \tilde I_a^2)^{1/2}} . \label{12b}
\end{eqnarray}
Note that $H_z(x,-\tilde I_a) \neq -H_z(x,\tilde I_a)$ only if
$R_{en}\neq R_{ex}$. Hence, the second harmonic $H_z^{(2)}(x)$
appears only when an edge barrier exists, and when this barrier
leads to an asymmetry with $I_{l} \neq I_{r}$. Using formulas
(\ref{2}), (\ref{3}), (\ref{11}), expressions (\ref{12a}),
(\ref{12b}) can be also rewritten in the form
 \begin{eqnarray}\label{12c}
\!\!H_z^{(1)}\!(x)\!\!\!&=&\!\!\!{1 \over 4\pi w}\!\!\left
(\![I_{a0}\!-\!I_e^{(1)}] \ln \Big |{w\!-\!x\over w\!+\!x}\Big
|\!+\!I_e^{(1)}\!\!{2wx\over w^2\!-\!x^2} \!\!\right )\!\!,~~ \\
\!\!H_z^{(2)}\!(x)\!\!\!&=&\!\!\!{w\over
2\pi(\!w^2\!-\!x^2\!)}\,\Delta  I_e^{(2)}, \label{12d}
 \end{eqnarray}
where $I_e^{(1)}$ is the first harmonic of $I_e$, while $\Delta
I_e^{(2)}$ is the second harmonic of $I_{r}-I_{l}$,
\begin{eqnarray}
I_e^{(1)}\!=\frac{4}{\pi}\int_0^1\!\![I_{r}(\tilde I_a\!)\!+
 \!I_{l}(\tilde I_a\!)]{\tilde I_a
 d\tilde I_a \over (1- \tilde I_a^2)^{1/2}}, \nonumber \\
\Delta I_e^{(2)}\!=\frac{4}{\pi}
 \int_0^1\!\![I_{r}(\tilde I_a\!)\!-
 \!I_{l}(\tilde I_a\!)] {(1-2\tilde I_a^2)
 d\tilde I_a \over (1- \tilde I_a^2)^{1/2}}. \nonumber
\end{eqnarray}

\subsection{General case}  

Formulas (\ref{4})-({\ref{10c}) have been obtained under the
assumption that the resistances are much larger than $\mu_0\omega$
(quasistatic approximation),
 \begin{equation}\label{13}
 R_{l},\ R_{r},\ R_b\ \gg \mu_0\omega ,
 \end{equation}
where $R_b=\rho(J/d)/2wd$. Under this assumption one can neglect
the inductance of the sample, which is of the order of $\mu_0$
(per unit length).\cite{ll} Besides this, when $R_b \gg
\mu_0\omega$, the two-dimensional penetration depth \cite{eh1,eh2}
$\Lambda=\rho_{\rm ff}/\mu_0\omega d$ of the ac field in the Ohmic
regime is considerably larger than the width $2w$ of the strip,
 \begin{equation}\label{13a}
{\Lambda \over 2w}={{\rm R}_{\rm ff}\over \mu_0\omega}\gg 1 ,
 \end{equation}
and hence, one may expect that the current distribution over the
cross section of the sample is indeed uniform. Since the
resistances $R_{l}$, $R_{r}$, $R_b$ decrease with decreasing
temperature $T$, the assumption (\ref{13}) is true for not too low
temperatures. We now address the general case when conditions
(\ref{13}) are not necessarily fulfilled. \cite{c1}

Let $E_a$ be the voltage drop (per unit length) which just
generates the current $I_a$. We now consider not only the edge
regions but formally also the whole strip as a set of parallel
wires. Then, we arrive at the following set of equations for the
left and right edge wires and for the wires of width $dx$ at
points $x$ ($-w<x<w$):
 \begin{eqnarray}\label{14a}
 E_a\!\!\!&=&\!\!\!R_lI_{l}\!+\!L_w \dot I_{l}\!+\!L_{lr}
 \dot I_{r}\!+\! \int_{-w}^{w}\!\!\!\!\!dx \dot J(x,t)L_{xl},\\
 E_a\!\!\!&=&\!\!\!R_rI_{r}\!+\!L_w \dot I_{r}\!+\!L_{lr}
 \dot I_{l}\!+\! \int_{-w}^{w}\!\!\!\!\!dx \dot J(x,t)L_{xr},
 \label{14b} \\
  E_a\!\!\!&=&\!\!\!\rho_x J(x)\!+\!L_{xl} \dot I_{l}\!+
  \!L_{xr}\dot I_{r}\!+\!\!\int_{-w}^{w}\!\!\!\!\!dx'
  \dot J(x'\!,t) L_{xx'}, \label{14c} \\
  I_a\!\!&=&\!\!I_{l}+I_{r}+I_b, \label{14d}
 \end{eqnarray}
where $\rho_x \equiv \rho[J(x)/d]/d$ is the sheet resistance,
\begin{equation}\label{14e}
 I_b=\int_{-w}^{w}\!\!\!\!\!dx J(x,t),
\end{equation}
the dot over currents means the time derivative,
$L_{xx'}=(\mu_0/2\pi)\ln (l/|x-x'|)$ is the mutual inductance per
unit length of two parallel wires of length $l$ located at points
$x$ and $x'$, \cite{ll} $L_{xl}=(\mu_0/2\pi)\ln (l/|x+w|)$ and
$L_{xr}=(\mu_0/2\pi)\ln (l/|x-w|)$ give the mutual inductances of
the edge wires and that located at $x$, $L_{lr}=(\mu_0/2\pi)\ln
(l/2w)$ is the mutual inductance of the two edge wires, and
$L_w=(\mu_0/2\pi)[\ln (l/r)+1/4]$ is the self inductance of the
edge wires. Here we have assumed that both edge wires have the
characteristic radius $r\sim d/2$. The magnetic field $H_z(x,t)$
is given by the Biot-Savart law
 \begin{eqnarray}\label{14f}  
 H_z(x,t)=\!{1 \over 2\pi }\!\left (
 \!{I_{r}\over w-x}-\!{I_{l}\over w+x}+
 \!\int_{-w}^w\!\!\! { J(x',t)\, dx' \over x'-x} \right )\!.
 \end{eqnarray}
Note that under conditions (\ref{13}), one may keep only the first
terms on the right hand sides of Eqs.~(\ref{14a})-(\ref{14c})
(omitting all time derivatives) and replace formula (\ref{14e}) by
Eq.~(\ref{4}) (since the skin effect is absent). After this
simplification, equations (\ref{14a})- (\ref{14d}) and (\ref{14f})
are equivalent to formulas (\ref{5}) - (\ref{11}).

In some region above temperature $T_d$ shown in Fig.~1, one has
$R_b \gg R_e$. Thus, a situation is possible in which $R_b \gg
\mu_0\omega > R_e$, and the skin effect is negligible, but one
cannot omit completely the terms with the inductances in
Eq.~(\ref{14a})-(\ref{14c}). In this context, it is useful to
consider the case when the only assumption is the uniform
distribution of the sheet current over $x$, i.e., the fulfilment
of Eq.~(\ref{4}). Under this assumption Eq.~(\ref{14a})-
(\ref{14d}) reduce to a form which describes three parallel
connected conductors with inductive coupling:
 \begin{eqnarray}\label{15a}
 E_a\!\!\!&=&\!\!\!R_lI_{l}\!+\!L_w \dot I_{l}\!+\!L_{lr}
 \dot I_{r}\!+\!L_{wb}\dot I_b,\\
 E_a\!\!\!&=&\!\!\!R_rI_{r}\!+\!L_w \dot I_{r}\!+\!L_{lr}
 \dot I_{l}\!+\!\!L_{wb}\dot I_b, \label{15b} \\
  E_a\!\!\!&=&\!\!\!R_bI_b\!+\!L_{wb} \dot I_{l}\!+
  \!L_{wb}\dot I_{r}\!+\!\!L_b \dot I_b, \label{15c} \\
  I_a\!\!&=&\!\!I_{l}+I_{r}+I_b, \label{15d}
 \end{eqnarray}
where $R_b=\rho(j)/2wd$ with $j=I_b/2wd$, $L_b=(\mu_0/2\pi)[\ln
(l/2w)+3/2]$ is the inductance of the strip for uniform current
distribution in it, and $L_{wb}=(\mu_0/2\pi)[\ln (l/2w)+1]$ is the
mutual inductance of the strip and one of the edge wires. In this
case Eq.~(\ref{14f}) transforms into expression (\ref{11}), and
formulas (\ref{12c}), (\ref{12d}) are valid.

Equations (\ref{14a})-(\ref{14d}) generalize the known equation
for a strip with time-dependent current, \cite{eh1,eh2} which
follows from Eq.~(\ref{14c}) if one omits the edge-wire terms with
$\dot I_{l}$ and $\dot I_{r}$. Thus, the numerical procedure
elaborated in Refs.~\onlinecite{eh1}, \onlinecite{eh2} can be
applied to solve Eqs.~(\ref{14a})- (\ref{14d}) in the general
case. However, in this paper we shall confine our analysis to
simple models which are sufficient to understand the main features
of the experimental results obtained in Refs.~\onlinecite{1,2,3}.

\section{Ohmic model}  

We begin our analysis with the simplified symmetric Ohmic model in
which $R_{en}=R_{ex}=2R_e$, and these resistances are independent
of the currents $I_{en}=I_{ex}$ but depend on temperature $T$. We
shall use a temperature dependence that satisfies the following
requirement: At $T=T_c$ when the edge barrier is absent, the
resistances are proportional to $R_b$ with a geometrical factor
that is the ratio of the cross-section area of the strip to the
cross-section area of one of the edge wires. At lower temperatures
we then take the conductance of the edge wires, $1/R_{en}$,
$1/R_{ex}$, as a sum of this bulk contribution and the conductance
caused by the edge barrier:
\begin{eqnarray}\label{16n}  
{1\over R_{en}(T)}\!=\!{1\over R_{ex}(T)}\! =\!{1 \over
R_0(T)}\!+\!{1 \over 2R_e^0}\!\left (\exp\!\!{{U_e(t)\over t}}
-1\right )\!,
\end{eqnarray}
where $R_0(T)=(w/r)R_b(T)$, $R_e^0$ is some constant, $t\equiv
T/T_c$, $T_c$ is the temperature of the superconducting
transition, $U_e(t)$ is the dimensionless magnitude of the edge
barrier (in units of $T_c$), and we shall imply a linear
temperature dependence of this barrier, $U_e(T)= U_e(0)(1-t)$,
with some constant $U_e(0)$. The function $R_0(T)$ just gives the
resistance of one of the edge wires with the edge barrier
neglected ($r\sim d/2$ is the characteristic radius of the edge
wire), while the second term in Eq.~(\ref{16n}) is the edge
contribution described by an Arrhenius law. We have introduced the
constant $-1$ in the brackets to take into account that this part
of the conductance vanishes at $T_c$. Besides this, in this Ohmic
model we use $j_c(T)=0$, and thus $R_b$ is independent of $J$ at
all temperatures. For definiteness we take the temperature
dependence of $R_b$ in the form of an Arrhenius law, too,
$R_b(T)=\rho_{\rm ff}/2wd=R_b(T_c)\exp[-U_b(0)(1-t)/t]$, but with
the constant $U_b(0)$ smaller than $U_e(0)$. The requirement
$U_b(0)<U_e(0)$ emphasizes that with decreasing temperature the
main contribution to the conductance of the edge wires results
from the edge barrier. In the discussion below, we shall imply
parameters that correspond to the experimental values: \cite{1,2}
$I_{a0}=4$ mA, $w=75\ \mu$m, $d=10\ \mu$m, $2w/d=15\ (\approx
w/r)$, $l=1.5\ $mm, $T_c=88$ K, $\omega/2\pi=73$ Hz,
$R_b(T_c)\approx 6.7\ \Omega/$cm, yielding the large value
$\Lambda/w=2.32\cdot 10^6$ for the two dimensional penetration
depth at $T=T_c$. For definiteness, in the construction of figures
we shall also take $R_e^0=R_0(T_c)$. As to $U_e(0)$ and $U_b(0)$,
we chose them so that one can reproduce the characteristic
temperatures $T_d$ and $T_{sb}$ marked in Fig.~1.

Within this simplified model, one has
$H_z^{(1)}(x)=H_z(x,t)/\sin\omega t$ and $H_z^{(2)}(x)=0$, where
under conditions (\ref{13}) (at sufficiently high temperatures)
$H_z(x,t)$ is given by Eq.~(\ref{11}), while $I_e (=2I_{l}
=2I_{r})$ and $I_b$ are described by Eqs.~(\ref{5}). Thus, we
obtain
 \begin{eqnarray}
 H_z^{(1)}\!(x)\!\!=\!\!{I_{a0} \over 4\pi w }\!\!
 \left (\!{R_e\over R_e\!+\!R_b} \ln\!\Big |{w\!-\!x\over w\!+
 \!x}\Big|\!+\!{R_b\over R_e\!+\!R_b}{2wx\over w^2\!-\!x^2}\!
  \right )\!\!.~ \label{16}
 \end{eqnarray}
With decreasing temperature when the inequality $R_e <\mu_0\omega$
becomes valid, it is necessary to take into account the
inductances of the strip and of the edge wires. In this case
Eqs.~(\ref{15a})-({\ref{15d}) reduce to a set of linear equations
(the time derivatives are replaced by the factor $i\omega$), and
after simple calculations we find
 \begin{equation}\label{17}
 2I_{l}=2I_{r}=I_e=
 I_a {\tilde z_b \over \tilde z_b+\tilde z_e}\,,\ \ \
 I_b=I_a {\tilde z_e \over \tilde z_b+\tilde z_e}\,,
 \end{equation}
where $\tilde z_b$ and $\tilde z_e$ are the ``effective
impedances'' of the strip and of the two edge wires,
 \begin{eqnarray}\label{18a}
\tilde z_b\!\!\!&=&\!\!R_b+i\omega(L_b-L_{wb})=R_b+i\omega{\mu_0
\over 4\pi}\,, \\
\tilde z_e\!\!\!&=&\!\!R_e\!+\!i\omega\! \left ({L_w\over
2}+\!{L_{lr}\over 2}-\!L_{wb}\right ) \nonumber \\
& &\ \ \ =R_e+i\omega{\mu_0 \over 4\pi} \left (\ln\!{2w \over r}
\,-{7\over 4}\right ). \label{18b}
 \end{eqnarray}
Note that even though the various inductances calculated per unit
length of the strip depend logarithmically on its length $l$, the
resulting effective impedances  $\tilde z_b$, $\tilde z_e$ (per
unit length) are independent of $l$. Inserting expressions
(\ref{17}) in formula (\ref{11}), one obtains $H_z^{(1)}(x)$. In
other words, down to the temperature $T_d$ defined in the Ohmic
model by $R_b(T_d)\sim \mu_0\omega$, the first harmonic is
described by formula (\ref{16}) in which $R_b$ and $R_e$ are
simply replaced by $\tilde z_b$ and $\tilde z_e$. The real part of
this expression gives the harmonic which is in-phase with $I_a$,
i.e., proportional to $\sin\omega t$ like $I_a(t)$, while the
imaginary part corresponds to the out-of-phase signal proportional
to $\cos\omega t$.

 \begin{figure}  
 \includegraphics[scale=.63]{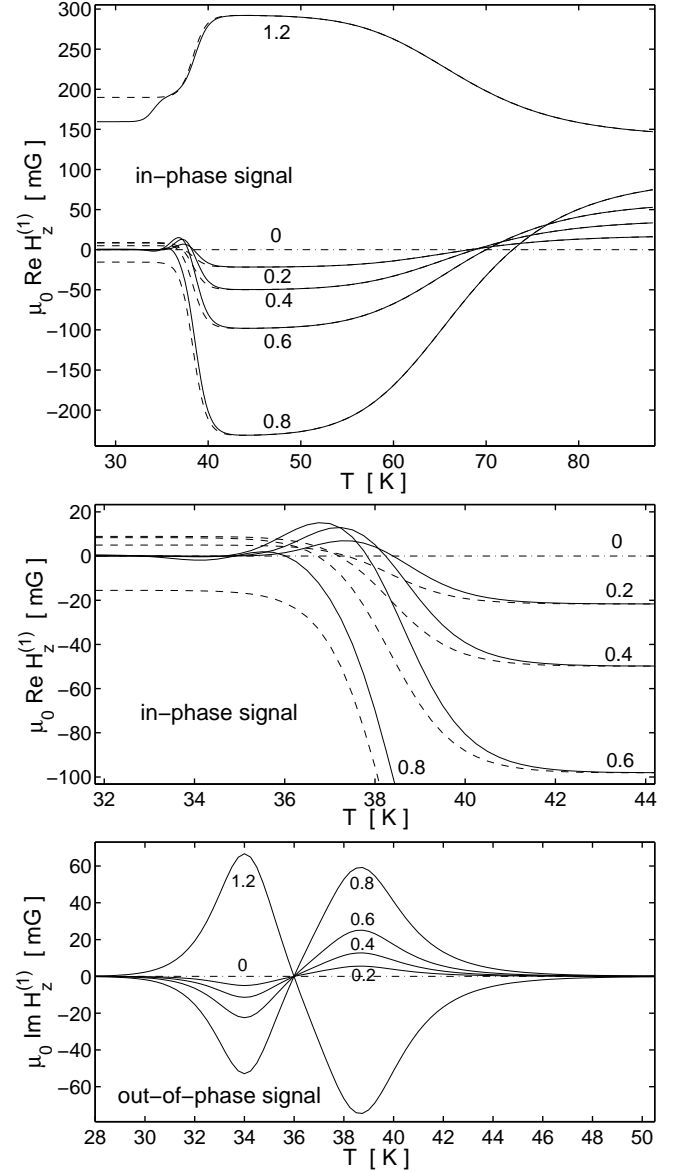}
\caption{\label{fig2} The temperature dependence of the first
harmonic of the ac self field which is in-phase with the applied
current, Re$H_z^{(1)}$, (top) and out-of-phase with it,
Im$H_z^{(1)}$, (bottom) at positions $-x/w=1.2$, 0.8, 0.6, 0.4,
0.2, 0 on the surface of a long thin strip of width $2w$. The
applied ac current is $I_a=I_{a0}\sin\omega t$. Results within the
symmetric Ohmic model with $R_{en}=R_{ex}$ for $U_e(0)=21$,
$U_b(0)=12$, $R_b(T_c)=6.7\ \Omega/$cm, $w/r=15$,
$R_e^0=15R_b(T_c)$, $\omega/2\pi=73$ Hz, $T_c=88$ K, $w=75\ \mu$m,
$I_{a0}=4$ mA. The dashed lines are the analytical results,
Eq.~(\ref{19}), while the solid lines are numerical results from
Eqs.~(\ref{14a})-(\ref{14f}) allowing for the skin effect. The
solid and dashed lines coincide at $T>42$ K, see the enlarged plot
(middle). Only numerical results are shown for Im$H_z^{(1)}$
(bottom) since the analytical results, Eq.~(\ref{19}), neglect the
skin effect and thus are applicable only at $T>42$ K. Note the
different scales of the $T$ axis for the different plots.
 } \end{figure}   

In Fig.~2 we show the temperature dependence of Re$H_z^{(1)}$ and
Im$H_z^{(1)}$ for discrete values of $x$ similar to the locations
of Hall sensors in the experiment. \cite{1,2} We first discuss the
temperature dependence of the in-phase signal Re$H_z^{(1)}$. At
temperatures near $T_c$, the bulk current $I_b$ exceeds $I_e$
since $R_b(T_c)<R_e(T_c)$, and one has an $H_z(x)$ that is
characteristic of the normal state of the superconductor [the
first term in Eq.~(\ref{16}) dominates], see Fig.~3. But the ratio
of the currents $I_e$ and $I_b$ changes with temperature. The role
of the edge current increases with decreasing $T$ since
$U_b(0)<U_e(0)$, and at a temperature $T=T_{sb}\approx 69$ K
defined by $R_e(T_{sb})=R_b(T_{sb})$ one has $I_b=I_e$, and the
contributions of $I_b$ and $I_e$ to $H_z$ practically compensate
each other for $x$ inside the strip. With further decrease of $T$,
the $x$-dependence of $H_z$ is mainly determined by the edge
current [i.e., by the second term in Eq.~(\ref{16})]. When the
inductances begin to play a role at temperature $T_x\approx 46$ K
defined by $R_e(T_x)\sim \mu_0\omega$ [or more exactly, by
Re$\tilde z_e(T_x) \sim {\rm Im}\tilde z_e(T_x)$], the impedance
$\tilde z_e$ saturates while $\tilde z_b$ still decreases with
decreasing $T$. As the temperature $T_d \approx 38$ K
corresponding to $R_b(T_d)\sim \mu_0\omega$ is approached, the
contribution of $I_b$ to $H_z$ begins to increase again, and the
ratio of the currents, $I_b/I_e$ tends to the constant
$\ln(2w/r)-7/4$, which is determined by the ratio of the imaginary
parts of $\tilde z_e$ and $\tilde z_b$. Of course, in reality at
such temperatures the skin effect plays an important role, and
Re$H_z^{(1)}$ tends to zero rather than to a profile determined by
the above-mentioned ratio of the currents, see Fig.~2. As to the
out-of-phase part of the first harmonic Im$H_z^{(1)}$, it
appears only near the temperature $T_x$ at which
the inductances begin to play a role, and it differs from zero
when the real and imaginary parts of the effective impedances are
comparable. At still lower temperatures when $R_e$ and $R_b$ become
negligible as compared to the imaginary parts of $\tilde z_e$ and
$\tilde z_b$, the out-of-phase signal vanishes again.

 \begin{figure}  
\epsfxsize= .9\hsize  \vskip 1.0\baselineskip \centerline{
\epsffile{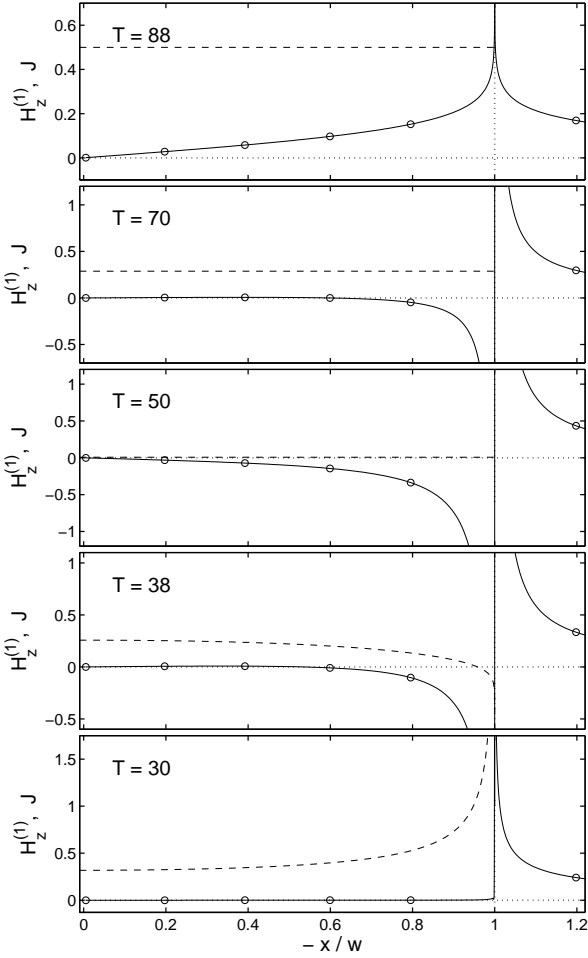}}  \vspace{.1cm}
\caption{\label{fig3} The in-phase components of the magnetic
field and of the sheet-current profiles, $H_z^{(1)}(x)$ (solid
lines) and $J(x)$ (dashed lines), for the symmetric Ohmic model
($\eta=0$) at temperatures $T=85$, $70$, $50$, $38$, and $30$ K,
accounting for the skin effect which occurs at $T<42$ K, cf.
Fig.~2. The parameters of the model are as in Fig.~2. The circles
mark the positions of the Hall probes used in Fig.~2. Both
$H_z^{(1)}$ and $J$ are in units of $I_{a0}/w$. The total current
in the depicted half strip is thus 1/2, equal to the area under
the dashed lines plus the (not shown) edge current at $x=w$.
 } \end{figure}   

We now present the results for the asymmetric case when
$R_{ex}\neq R_{en}$. In this situation model (\ref{16n}) is
generalized as follows:
\begin{eqnarray}  
{1\over R_{ex}(T)}={1 \over R_0(T)}+{1 \over R_{ex}^0}\left
(\exp \!{U_e(t)\over t} -1\right ), \nonumber \\
{1\over R_{en}(T)}={1 \over R_0(T)}+{1 \over R_{en}^0}\left (\exp
\!{U_e(t)\over t} -1\right ), \label{24b}
\end{eqnarray}
where $R_{ex}^0(T)$, $R_{en}^0(T)$ are some constants, and
$R_0(T)$ is the same as in Eq.~(\ref{16n}). It is convenient to
introduce the parameter
\begin{equation}\label{24a}
 \eta\equiv {R_{ex}^0-R_{en}^0\over R_{ex}^0+R_{en}^0}
 \end{equation}
which characterizes the asymmetry of the edge barrier. Then, one
has
\[
 R_{en}^0=2R_e^0/(1+\eta), \ \ \ R_{ex}^0=2R_e^0/(1-\eta),
 \]
where $R_e^0=R_{en}^0R_{ex}^0/(R_{en}^0+R_{ex}^0)$. We emphasize
that the constant $\eta$ defines the ratio $(R_{ex}-R_{en})/
(R_{ex}+R_{en})$ at sufficiently low temperatures when the
conductance $1/R_0$ is negligible in Eqs.~(\ref{24b}), while at
$T\to T_c$ the resistances $R_{en}$ and $R_{ex}$ tend to $R_0$,
the resistance of the edge wires without barrier. In other words,
formulas (\ref{24b}) describe both the asymmetry of the edge wires
at low temperatures, and the disappearance of the asymmetry when
the edge barrier vanishes. Note also that $R_e= R_{ex}R_{en} /
(R_{ex}+R_{en})$ is not equal to $R_e^0$ but approaches it with
decreasing temperature.

In this asymmetric case the first harmonic becomes:
 \begin{eqnarray}
H_z^{(1)}\!(x)\!=\!{I_{a0} \over 4\pi w }\!\left (
 \!{\tilde z_{e\eta} \over \tilde z_{e\eta}\!+\!\tilde z_b}
 \ln\Big |{w\!-\!x\over w\!+\!x}\Big|\!+\!{\tilde z_b\over
 \tilde z_{e\eta}\!+\!\tilde z_b}{2wx\over w^2\!-\!x^2}\!
  \right )\!\!,~ \label{19}
 \end{eqnarray}
where
\begin{eqnarray}\label{20}
 \tilde z_{e\eta}&=&\tilde z_e+ {(R_{ex}-R_{en})^2 \over
 4(R_{ex}+R_{en})} {i p\, \over 1+ip}\,,
 \end{eqnarray}
$\tilde z_e$, $\tilde z_b$ are given by formulas (\ref{18a}),
(\ref{18b}), and
 \begin{equation}\label{21}
 p\equiv {2\omega(L_w-L_{lr})\over R_{ex}+R_{en}}=
 {\mu_0\omega \over \pi (R_{ex}+R_{en})}
 \left (\ln\!{2w \over r}+{1\over 4}\right ).
 \end{equation}
This $p$ characterizes the relative contribution of the inductance
of the edge wires to their effective impedance $\tilde z_e$.
Analysis of formulas (\ref{19})-(\ref{21}) shows that the
parameter $\eta$ practically has no effect on $H_z^{(1)}$ within
the considered Ohmic model, and hence the first harmonic in the
asymmetric case, in fact, coincides with the harmonic in the
symmetric case of $\eta=0$. However, for the case $\eta \neq 0$
the second harmonic $H_z^{(2)}$ appears. This $H_z^{(2)}(x)$,
Eq.~(\ref{12d}), is
\begin{equation}\label{22}
 H_z^{(2)}\!(x)\!\!=\!\!{2I_{a0}\over 3\pi^2}
 {w \over w^2\!-\! x^2}{(R_{ex}\!-\!R_{en})\over (R_{ex}\!+\!R_{en})}
 {\tilde z_b \over \tilde z_b+\tilde z_{e\eta}}{1\over 1+ip}.
\end{equation}
The real part of this expression gives the harmonic proportional
to $\cos2\omega t$, and its imaginary part to
$\sin2\omega t$. In Fig.~4 we show the temperature dependence of
$|H_z^{(2)}|$ at $x=0.7w$. At temperatures near $T_x$ introduced
by Re$\tilde z_e(T_x) \sim {\rm Im}\tilde z_e(T_x)$, one has
$p(T_x)\sim 1$, the amplitude of the second harmonic begins to
decrease with decreasing $T$ and becomes small even at
temperatures above $T_d$ if $R_b(T_x)\gg \mu_0\omega$. Of course,
at temperatures when $R_b(T)\sim \mu_0\omega$, the second harmonic
$H_z^{(2)}$ vanishes in any case due to the skin effect.

 \begin{figure}  
 \includegraphics[scale=.46]{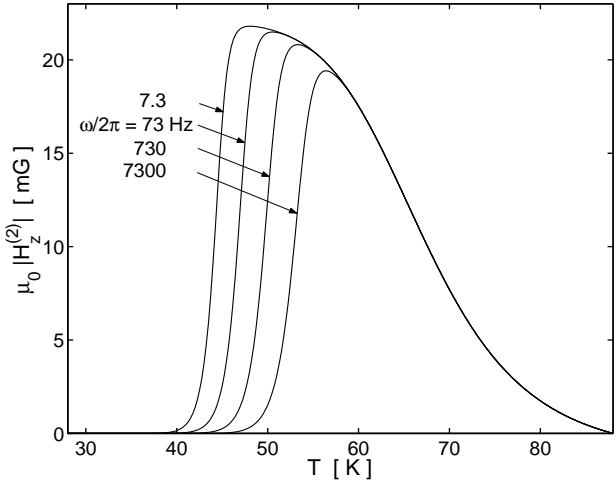}
\caption{\label{fig4} The temperature dependence of the magnitude
of the second harmonic of the ac self field, $|H_z^{(2)}|$, at
position $|x/w|=0.7$ on the surface of a long thin strip of width
$2w$ with applied ac current $I_a=I_{a0}\sin\omega t$. Results
within the Ohmic model, Eq.~(\ref{22}), for the same parameters as
in Fig.~2 but with asymmetry $\eta=0.1$ and amplitude $I_{a0}=10$
mA (line at $\omega/2\pi=73$ Hz). The other lines are for
different frequencies.
 } \end{figure}   

We emphasize that the formulas of this section are valid not only
for the model dependences (\ref{16n}) or (\ref{24b}) used here but
also for any functions $R_b(T)$, $R_{en}(T)$, $R_{ex}(T)$. Thus,
one can obtain some information on these functions fitting the
theoretical dependences of $H_z^{(1)}(T)$ and $H_z^{(2)}(T)$ to
appropriate experimental data. Independent information on the
functions $R_b(T)$, $R_e(T)$ for each specific sample can be also
obtained from usual transport measurements. \cite{4} Beside this,
some data on this subject can be extracted from frequency
dependences of $H_z^{(1)}(T)$ and $H_z^{(2)}(T)$ since these
harmonics are functions of the combinations $(R_b/\mu_0\omega)$,
$(R_{en}/\mu_0\omega)$, $(R_{ex}/\mu_0\omega)$. In the framework
of our model (\ref{24b}), if one changes the frequency $\omega$
from $\omega_1$ to $\omega_2$, the temperature dependence of the
first harmonic changes only near $T_d$, while the temperature
dependence of the second harmonic changes in the vicinity of $T_x$
(the appropriate parts of these dependences shift to higher
temperatures with increasing $\omega$), see, e.g., Fig.~4. The
shifts of the temperatures $T_d$ and $T_x$ are
 \begin{eqnarray} \label{te}
 T_d(\omega_2)-T_d(\omega_1)\approx {T_d^2\over U_b(0)T_c}
 \ln\left ({\omega_2 \over \omega_1}\right ), \nonumber \\
 T_x(\omega_2)- T_x(\omega_1)\approx {T_x^2 \over U_e(0)T_c}
 \ln\left ({\omega_2 \over \omega_1}\right ),
 \end{eqnarray}
and they give information on $U_b(0)$ and $U_e(0)$.

We now briefly discuss the applicability of the considered
temperature dependences of $R_e$ and $R_b$ to BSCCO samples. The
experimental data on resistance of BSCCO strips obtained from
usual transport measurements are well approximated by an Arrhenius
law. \cite{P,P1,B,4} In principle, parameters $U_e(0)$ and $R_e^0$
can be found from these data, while $U_b(0)$ can be estimated in
similar experiments with wide samples.\cite{4} In particular, in
Ref.~\onlinecite{4} it was found that $U_e(0)\approx 12$ for the
same crystal as in Fig.~2, and it was estimated that $U_b(0)\le
8$. Both these values are noticeably less than those used in the
construction of Figs.~2 and 4. The latter values were chosen such
that one can reproduce the experimental $T_d$ and $T_{sb}$. In
Figs.~2 and 4 we have also used the specific value of $R_e^0$,
$R_e^0=R_0(T_c)$. Other choices of the parameter $R_e^0$ enable
one, in principle, to use a smaller value of $U_e(0)-U_b(0)$, but
do not provide the coincidence of both $U_e(0)$ and $U_b(0)$ with
their experimental values [in the Ohmic model the value of
$U_b(0)\approx 12$ is fixed by the condition $R_b(T_d)\approx
\mu_0\omega$]. \cite{c} Such disagreement seems to signal that the
Ohmic model does not work in the whole interval from $T_d$ to
$T_c$ for the BSCCO crystals.

Finally, we note two characteristic features of the Ohmic model
studied here. First, if the magnetic field is measured in units of
$I_{a0}/w$, the first and the second harmonics are independent of
the amplitude $I_{a0}$ of the applied current. In particular, the
characteristic temperatures $T_d$, $T_{bs}$, $T_x$ do not shift
with a change of $I_{a0}$. This scaling property may be used in
experiments to find the temperature regions where the Ohmic model
is applicable. Second, the resistance of the strip,
$R=R_bR_e/(R_b+R_e)$ is also independent of $I_{a0}$ and can be
used for the same aim as well.

\section{Model of nonlinear $R_e(I_e)$}  

The data of Ref.~\onlinecite{3} obtained for different currents
$I_{a0}$ show that at least for some superconductors $R_e$ cannot
be considered as a quantity that is independent of $I_e$ for all
temperatures. To get an insight into this problem, we shall now
analyze a simple model of nonlinear dependence $R_e(I_e)$ and show
that this nonlinearity can provide an alternative description of
the experimental temperature dependences of $H_z^{(1)}$  and
$H_z^{(2)}$ near $T_{sb}$. In principle, this and a similar
nonlinear model for $R_b(I_b)$ near $T_d$ enable one to describe
the experimental data on $H_z^{(1)}$ and $H_z^{(2)}$ with
realistic values of $U_e(0)$ and $U_b(0)$.

We shall consider the following simple model dependences
$R_{en}(I_{en})$ and $R_{ex}(I_{ex})$ for the flux entrance into
the strip and the flux exit from it: When $I_{en}$ (or $I_{ex}$)
exceeds the critical current $I_{en}^c$ ($I_{ex}^c$) for flux
entrance (exit), the resistivity of the edge wire is the same as
in the bulk of the strip, and hence the resistance of the edge
wire $R_{en}(I_{en})$ [$R_{ex}(I_{ex})$] coincides with
$R_0=(w/r)R_b$. At $I_{en}=I_{en}^c$ ($I_{ex}=I_{ex}^c$) the
resistance sharply drops down to a value $R_{en}$ ($R_{ex}$) which
is independent of the current $I_{en}$ ($I_{ex}$) for $I_{en}
<I_{en}^c$ ($I_{en}< I_{en}^c$), i.e., one has
\begin{eqnarray}\label{23a}
R_{en}(I_{en})=R_0 \ \ {\rm for}\ \ I_{en}>I_{en}^c, \nonumber \\
R_{en}(I_{en})=R_{en} \ \ {\rm for}\ \ I_{en}<I_{en}^c,
\end{eqnarray}
and
\begin{eqnarray}\label{23b}
R_{ex}(I_{ex})=R_0 \ \ {\rm for}\ \ I_{ex}>I_{ex}^c, \nonumber \\
R_{ex}(I_{ex})=R_{ex} \ \ {\rm for}\ \ I_{ex}<I_{ex}^c.
\end{eqnarray}
In other words, we replace the complicated nonlinear dependence of
the edge barrier on the current flowing in the edge region
\cite{bkv} by a sharp jump of the appropriate resistance. Although
with the equations of Sec.~II, the currents and magnetic field can
be found for any realistic dependences $R_{en}(I_{en})$ and
$R_{ex}(I_{ex})$, within this simple model the calculations can be
carried out analytically. This enables one to get insight into
situations with various relations between the edge critical
currents $I_{en}^c$, $I_{en}^c$ and the resistances $R_{en}$,
$R_{ex}$.

 \begin{figure}  
 \includegraphics[scale=.48]{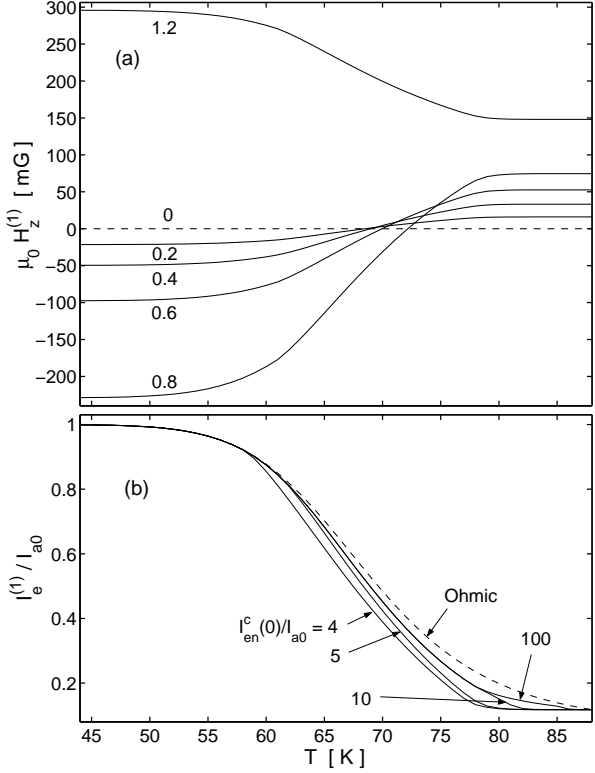}
\caption{\label{fig5}Top: Temperature dependence of the first
harmonic $H_z^{(1)}$ of the ac self field at positions $x/w=-1.2$,
$-0.8$, $-0.6$, $-0.4$, $-0.2$, $0$ on the surface of a long thin
strip of width $2w$ with applied ac current $I_a=I_{a0}\sin\omega
t$. Results within the nonlinear $R_e$-model for the same
parameters as in Fig.~2 (symmetric case, $\eta=0$) but with
$U_e(0)=12$, $U_b(0)=2$, $I_{en}^c(0)/I_{a0}= I_{ex}^c(0)/I_{a0}
=4$. Assumed was $I_{en}^c(T)/I_{en}^c(0) =I_{ex}^c(T)/
I_{ex}^c(0) =(1-t)^2$ where $t\equiv T/T_c$. Bottom: Temperature
dependence of the first harmonic $I_e^{(1)}$ of the edge current.
The parameters are the same as in the top panel but
$I_{en}^c(0)/I_{a0}=4$ (symmetric case), $5$, $10$, $100$. If
$\eta$ differs from zero, the curves remain practically unchanged
for $\eta < 0.3$. For comparison, the dashed line shows
$I_e^{(1)}(T)$ for the symmetric Ohmic model with parameters of
Fig.~2.
 } \end{figure}   

Let us specify the details of this model. With decreasing
temperature the currents $I_{en}^c$ and $I_{ex}^c$ increase, and
for definiteness we take $I_{en}^c=I_{en}^c(0)(1-t)^2$,
$I_{ex}^c=I_{ex}^c(0)(1-t)^2$ where $t\equiv T/T_c$ (such
temperature dependence of $I_{en}^c+I_{ex}^c$  was obtained in
Ref.~\onlinecite{bkv}). We also imply that near the temperature
$T_{sb}$ these critical currents of the edge wires become larger
than the amplitude of the applied current $I_{a0}$, and thus the
drop of edge-wire resistances occurs at sufficiently high
temperatures where conditions (\ref{13}) are true, and so the
inductances and the skin effect are ignored below. As to $R_{en}$,
$R_{ex}$ and $R_b$,  we choose the same expressions for them as in
the Ohmic model, but now we use $U_e(0)\approx 12$ and
$U_b(0)\approx 2$ (in units of $T_c$), which are essentially less
than those used in Sec.~III.

To find the first and second harmonics of $H_z$, one should solve
Eqs.~(\ref{10a})-(\ref{10c}) and find the currents $I_{l}$,
$I_{r}$, and $I_b$ as functions of $I_a$. From these currents, one
calculates $H_z^{(1)}$, $H_z^{(2)}$ using formulas
(\ref{12c})-(\ref{12d}). Within the above simple model, these
calculations can be done analytically. For example, in the
symmetric case when $\eta=0$ and $I_{en}^c=I_{ex}^c\equiv
I_e^c/2$, one has formulas (\ref{5}) for $I_b$ and
$I_e=2I_{l}=2I_{r}$ at $I_a \le I_1\equiv I_e^c\cdot
(R_b+R_e)/R_b$ where $R_b$ and $R_e=R_{en}R_{ex}/(R_{en}+R_{ex})$
are independent of $I_b$ and $I_e$. In the interval $I_1\le I_a
\le I_2\equiv I_e^c\cdot (1+w/2r)$ we find $I_e=I_e^c$, $I_b=I_a-
I_e^c$, while at $I_2\le I_a$ equations (\ref{10a})-(\ref{10c})
yield
\begin{eqnarray}\label{24}
 I_e&=&I_a {R_b\over R_b+0.5R_0}=I_a\cdot \left (1+{w\over 2r}\right
 )^{-1}, \nonumber \\
 I_b&=&I_a {0.5R_0\over R_b+0.5R_0}=I_a {w\over 2r}
 \left (1+{w\over 2r}\right )^{-1}.
\end{eqnarray}
Using these formulas, we find the following expression for the
first harmonic of $I_e$:
\begin{eqnarray}\label{25}
 I_e^{(1)}\!\!\!\!&=&\!\!\!\frac{2I_{a0}}{\pi}{R_b\over
 R_b+R_e}\!\left (\!\phi_1\!-\!{\sin2\phi_1\over 2}\!\right)\!
 +\!\frac{4I_e^c}{\pi}(\cos\phi_1 \nonumber \\
&-&\!\!\cos\phi_2)\!+\!\frac{2I_{a0}}{\pi}\!\left (\frac{\pi}{2}\!
-\!\phi_2\!+\!{\sin2\phi_2\over 2}\!\right)\!\!\left (1+{w\over
2r}\right )^{\!-1}\!\!\!\!\!\!,~~~~
\end{eqnarray}
where $\phi_1\equiv \min({\rm arcsin}[I_1/I_{a0}],\pi/2)$ and
$\phi_2\equiv \min({\rm arcsin}[I_2/I_{a0}],\pi/2)$. Inserting
Eq.~(\ref{25}) into formula (\ref{12c}), we obtain $H_z^{(1)}(x)$.
The second harmonic of $H_z$ in the symmetric case is equal to
zero. In Appendix A we present the first and the second harmonics
of $H_z$ in the asymmetric case when $\eta\neq 0$, $I_{en}^c\neq
I_{ex}^c$. Note that within this nonlinear $R_e$ model the
characteristic temperature $T_{sb}$ is still defined by the
condition $I_e^{(1)}= I_{a0}/2$ as in the Ohmic model.

In Fig.~5 we show the first harmonics $H_z^{(1)}$ of $H_z(t)$
calculated within this nonlinear $R_e$ model. The temperature
dependence of the first harmonic is qualitatively similar to that
of Fig.~2, and with increasing asymmetry of the edge barrier it
does not change essentially. On the other hand, the second
harmonic $H_z^{(2)}$ is highly sensitive to the asymmetry, see
Fig.~6. Figure 6 shows both the change of $H_z^{(2)}$ with
$I_{en}^c(0)/I_{ex}^c(0)$ at fixed $\eta$ and its change with
$\eta$ at fixed $I_{en}^c(0)/I_{ex}^c(0)$. Interestingly, the
temperature dependence of $H_z^{(2)}$ is nonmonotonic and looks
qualitatively similar to that of Fig.~4. But now the temperature
where $H_z^{(2)}(T)$ sharply decreases is due to the temperature
$T_1$ of Appendix A at which some characteristic current of the
edge barrier becomes larger than $I_{a0}$. This temperature $T_1$
is independent of $\omega$ within our nonlinear $R_e$ model. At
temperatures $T\le T_1$ during the whole period of the ac field,
the currents flowing in the edge wires are less than the
appropriate critical currents $I_{en}^c$, $I_{ex}^c$, the strip is
in the Ohmic regime, and $H_z^{(1)}(T)$ and $H_z^{(2)}(T)$
obtained from the formulas of Appendix A coincide with expressions
(\ref{19}) and (\ref{22}) if one neglects the inductive terms
proportional to $\mu_0\omega$ in these expressions. This
coincidence permits one to combine this nonlinear model with the
Ohmic model of Sec.~III and thus to obtain continuous curves in
the entire temperature region. Interestingly, when $T$ is larger
than the temperature $T_4$ of Appendix A, the maximum currents in
the edge wires exceed $I_{en}^c$ and $I_{ex}^c$, and the strip is
in the flux-flow regime during some part of the ac period. In
Fig.~6 this temperature $T_4$ corresponds to the point where
$H_z^{(2)}(T)$ crosses zero (i.e., $T_4\approx 78$ K in Fig.~6b).

 \begin{figure}  
 \includegraphics[scale=.48]{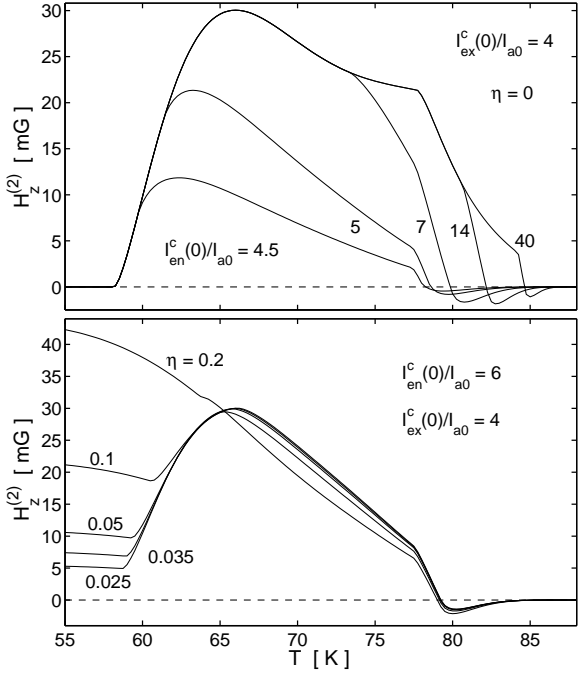}
\caption{\label{fig6} The temperature dependence of the second
harmonic of the ac field defined by Eq.~(\ref{1b}) at $|x/w|=0.7$
within the nonlinear $R_e$ model, Eqs.~(\ref{12d}), (\ref{A4}).
The parameters are the same as in Fig.~2 but with $U_e(0)=12$,
$U_b(0)=2$, and besides this $I_{ex}^c(0)/I_{a0}=4$. The top part
shows the second harmonic at fixed $\eta=0$ for different
$I_{en}^c(0)/I_{a0}=4.5$, $5$, $7$, $14$, $40$, while the bottom
part gives this harmonic at fixed $I_{en}^c(0)/I_{a0}=6$ but for
different asymmetry $\eta=0.025$, $0.035$, $0.05$, $0.1$, $0.2$.
Note that $H_z^{(2)}(T)$ for $\eta \neq 0$ does not reach zero
after its drop in the vicinity of $60$ K.
 } \end{figure}   

\section{Model of nonlinear $R_b(I_b)$}  

As was mentioned above, in the Ohmic model below a temperature
$T_d$ defined by the condition $R_b(T_d)\sim \mu_0\omega$, the
skin effect occurs, which means the perpendicular ac magnetic
field $H_z$ is expelled from the strip. However, there is another
reason for this field to be expelled from the superconductor.
Indeed, if the ac current flowing in the bulk is smaller than the
critical current $I_c=2wdj_c$ of the strip, the ac field cannot
penetrate completely into the sample. Thus, a finite $I_c$ and its
increase with decreasing temperature can, in principle, explain
the existence of the temperature $T_d$ in Fig.~1, below which
$H_z$ is zero in the strip and at its surface. To describe this,
we now consider a nonlinear $R_b$ model in which the critical
current density $j_c$ is not equal to zero, and the resistivity
$\rho$ depends on the current density $j$. We shall analyze the
following model current--voltage dependence:
\begin{equation}\label{51}   
E=\rho(j)j=\rho_{\rm ff}{j j_c \epsilon(j) \over j+j_c
\epsilon(j)},
\end{equation}
where
\begin{equation}\label{52}
 \epsilon(j)=2 e^{-U_0/T}\sinh \left ({j\over j_c} {U_0\over
T}\right ),
\end{equation}
and $j_c$ and the effective depth of flux-pinning well $U_0$ are
some decreasing functions of $T$. Let us consider the case $U_0/T
\gg 1$. Then, at $j>j_c$ one has $\epsilon(j)\gg j/j_c$, and thus
$\rho(j)=\rho_{\rm ff}$, while at $j<j_c$ the quantity
$\epsilon(j)$ is small, $\epsilon(j)\ll j/j_c$, and we arrive at
$\rho(j)j=\rho_{\rm ff}j_c\epsilon(j)\ll \rho_{\rm ff}j$.\cite{c2}
Thus, equations (\ref{51}), (\ref{52}) lead to a sharp jump of
resistivity at $j=j_c$ that is characteristic for the critical
state model. In analytical calculations we shall simplify the
model dependence further, putting
\begin{eqnarray}\label{53}
\rho(j)&=&\rho_{\rm ff} \ \ \ \ \ {\rm for}\ j>j_c, \nonumber \\
\rho(j)&=&0\ \ \ \ \ \ \ {\rm for}\ j<j_c.
\end{eqnarray}
As to the flux-flow resistivity $\rho_{\rm ff}$, for definiteness
we shall imply the same temperature dependence as in Secs.~III and
IV, $\rho_{\rm ff}(T)=\rho_{\rm ff}(T_c)\exp[-U_b(0)(1-t)/t]$,
where $U_b(0)$ is a dimensionless constant and $t\equiv T/T_c$.
Since in this section we are mainly interested in understanding
the effect of nonlinear $R_b(I_b)$ on the first harmonic of the ac
field near $T_d$, we assume that near $T_d$ one has $R_{\rm
ff}=\rho_{\rm ff}/2wd \gg \mu_0\omega$, $R_e$, and the resistances
$R_{en}=R_{ex}=2R_e$ are independent of $I_{en}=I_{ex}$ (i.e., we
treat only the symmetric situation and imply that the skin effect
occurs sufficiently below $T_d$). The temperature dependence of
the critical current $I_c=2wdj_c$ is assumed to be a monotonically
decreasing function. From our analysis it will be clear that this
dependence plays the most important role in the vicinity of the
temperature $T_j$ at which $I_c(T)$ reaches $I_{a0}$, and so we
may use the following representation of $I_c(T)$:
\begin{equation}\label{59a}
I_c(T)= I_{a0}\left [T_c-T\over T_c-T_j\right ]^{n},
\end{equation}
where $n$ is some exponent. Note that in this representation a
change of $I_{a0}$ requires a change of $T_j$ such that
$I_c(T)$ remains unchanged.

 \begin{figure}  
 \includegraphics[scale=.48]{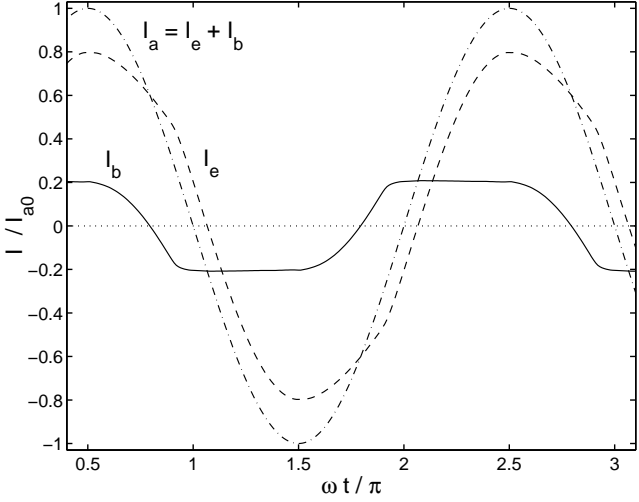}
\caption{\label{fig7} The temporal dependences of the applied
current $I_a=I_{a0}\sin\omega t$, the bulk current $I_b$, and the
edge current $I_e$ obtained numerically from
Eqs.~(\ref{14a})-(\ref{14e}) with the realistic current-voltage
dependence (\ref{51}), (\ref{52}) at $T=36.5$ K and $U_0/T=30$.
The other parameters are the same as in Fig.~2 but $U_b(0)=6$,
$U_e(0)=15$, and $I_{a0}=3.57I_c$. At this temperature formula
(\ref{56}) gives $I_2/I_c\approx 3700$. With increasing $I_{a0}$
the width of time intervals where $I_b$ changes from $I_c$ to
$-I_c$ or vice versa decreases.
 } \end{figure}   

 \begin{figure}  
 \includegraphics[scale=.48]{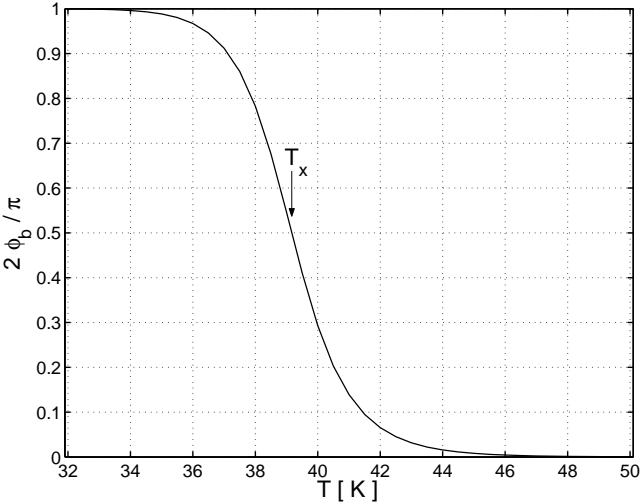}
\caption{\label{Fig8} The temperature dependence of the phase
shift $\varphi_b$ of the bulk current $I_b(t)$ as compared to the
applied current $I_a(t)=I_{a0}\sin\omega t$, Eq.~(\ref{55}). The
parameters are the same as in Fig.~2 but $U_b(0)=6$, $U_e(0)=15$.
The temperature $T_x \approx 39$ K is defined by Re$\tilde
z_e(T_x)={\rm Im}\tilde z_e(T_x)$, i.e., by $\varphi_b\approx
\pi/4$.
 } \end{figure}   

Let us first show that at $I_{a0}\gg I_c=2wdj_c$ our nonlinear
$R_b$ model reduces to the Ohmic model considered in Sec.~III.
This situation always occurs at sufficiently high temperatures
when $I_c$ is small. The nonlinearity of $R_b(I_b)$ gives only
small corrections to the Ohmic results, and these corrections can
be calculated analytically. Indeed, at times $t$ when
$I_b(t)>I_c$, the Ohmic model is applicable, and we have
 \begin{equation}\label{54a}
I_b=I_{b0}\sin(\omega t+\varphi_b),\ \ \  I_e=I_{e0}\sin(\omega
t+\varphi_e)
 \end{equation}
where $I_{b0}$ and $I_{e0}$ are the amplitudes of the currents
$I_b$ and $I_e$,
\begin{eqnarray}\label{54}
I_{b0}&=&I_{a0}{|\tilde z_e|\over |R_{\rm ff}+\tilde z_e|}\approx
I_{a0}{|\tilde z_e|\over R_{\rm ff}}, \nonumber \\
I_{e0}&=&I_{a0}{R_{\rm ff} \over |R_{\rm ff}+\tilde z_e|}\approx
I_{a0},
\end{eqnarray}
the effective impedance of the wires $\tilde z_e$ is given by
Eq.~(\ref{18b}), and $\varphi_b$, $\varphi_e$ define the phase
shifts of $I_b$ and $I_e$ with respect to $I_a=I_{a0}\sin\omega
t$,
\begin{equation}\label{55}
\tan \varphi_b\approx {{\rm Im}\tilde z_e\over {\rm Re}\tilde
z_e},\ \ \ \tan \varphi_e\approx -{{\rm Im}\tilde z_e \over R_{\rm
ff}} \ll 1.
\end{equation}
Formulas (\ref{54a}) - (\ref{55}) are valid for those times $t$
when $|I_b(t)|$ defined by these formulas exceeds $I_c$. When the
absolute value of $I_b(t)$ reaches $I_c$, the bulk current $I_b$
remains equal to $\pm I_c$, while $I_e=I_a-I_b$, see
Fig.~\ref{fig7}. \cite{c3} It is important that $I_b$ is uniformly
distributed over the width of the strip, $J(x)=\pm I_c/2w=\pm
j_cd$, so long as $|I_b|=I_c$. This is due to the sharpness of the
current-voltage law (\ref{53}), which prevents the skin effect. A
nonuniform distribution of the sheet current $J$ over $x$ occurs
only in a relatively narrow time interval (as compared to the
period $T=2\pi/\omega$) when $I_b$ changes from $I_c$ to $-I_c$ or
vice versa. The relative width of this interval is determined by
the small ratio $I_c/I_{a0}\ll 1$, and in the first approximation
we can neglect this time interval in calculating the first
harmonic of $I_b(t)$,
\begin{equation}\label{57}
I_b^{(1)}\equiv \frac{2}{T}\int_0^T I_b (t)\sin{\omega t}dt,
\end{equation}
which is in-phase with the applied current. Then, we find
\begin{eqnarray}\label{58}
I_b^{(1)}\!\!\!\approx\!\!{2I_{b0}\over
\pi}\cos\varphi_b\!\cdot\!\! \Big [ {\pi\over 2}\!-\!
\varphi_2\!+\! {\sin 2\varphi_2 \over 2}\!
 +\!{2I_2 \over I_{a0}}(1\!-\cos\varphi_2)\Big ]\!,~
\end{eqnarray}
where $\varphi_2\equiv \min({\rm arcsin}[I_2/I_{a0}],\pi/2)$,
$I_2$ is the total current at which the amplitude of the bulk
current, $I_{b0}$, becomes equal to $I_c$,
\begin{equation}\label{56}
I_2\equiv I_c {|R_{\rm ff}+ \tilde z_e|\over |\tilde z_e|} \approx
I_c {R_{\rm ff}\over |\tilde z_e|},
\end{equation}
and $I_{b0}$, $\varphi_b$ are determined by formulas (\ref{54}),
(\ref{55}). If $I_{a0}\gg I_2$, one has $\varphi_2\ll 1$, and
formula (\ref{58}) indeed reduces to the result for the Ohmic
model, $I_b^{(1)}\approx I_{b0}\cos\varphi_b$. On the other hand,
at $I_2>I_{a0}\gg I_c$ we have $\varphi_2=\pi/2$, and this formula
gives
\begin{equation}\label{59}
I_b^{(1)}\approx {4\over \pi}I_c \cos\varphi_b.
\end{equation}
Thus, in this case the temperature dependence of $I_b^{(1)}$ is
straightforwardly expressed via the temperature dependence of the
critical current and of the phase shift $\varphi_b$. The
dependence $\varphi_b(T)$ is shown in Fig.~8. Taking into account
that $I_e^{(1)}= I_{a0}- I_b^{(1)}$ and using formula (\ref{12c}),
which is always valid in the absence of the skin effect, we obtain
the first harmonic of the magnetic field under condition
$I_{a0}\gg I_c$, i.e., when the calculated $H_z^{(1)}(T)$ does not
deviate considerably from the appropriate result of the Ohmic
model, see Fig.~9. Here $H_z^{(1)}$ is the first harmonic defined
by Eq.~(\ref{1a}), i.e., its in-phase part.

 \begin{figure}  
 \includegraphics[scale=.48]{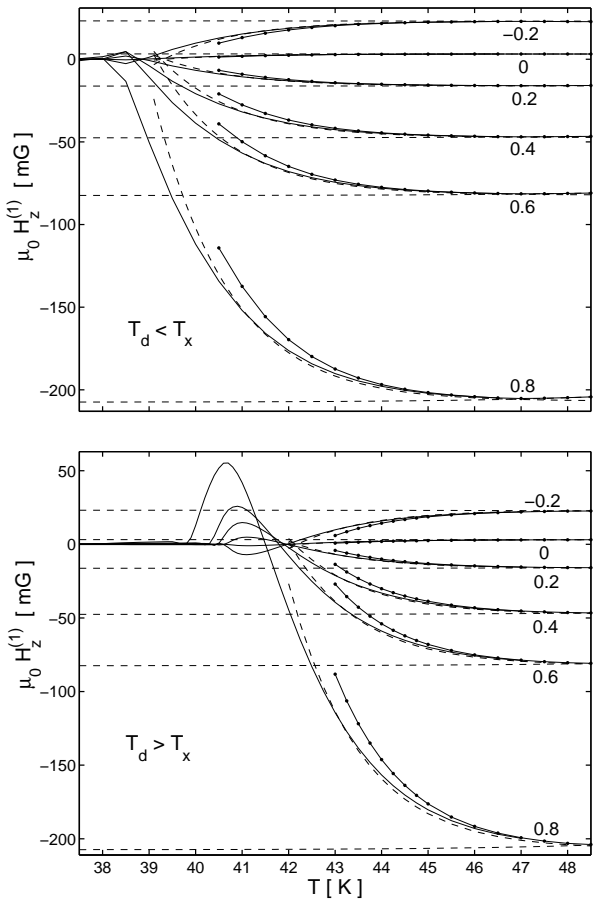}
\caption{\label{fig9} The temperature dependences of the first
harmonic $H_z^{(1)}$ defined by Eq.~(\ref{1a}) within the
nonlinear $R_b$ model. The parameters are the same as in Fig.~2
but $U_b(0)=6$, $U_e(0)=15$. The critical current $I_c$ is
modelled by the power law $I_c/I_{a0}=[(T_c-T)/(T_c-T_j)]^{30}$
with $T_j=37$ K (top) and $T_j=40$ K (bottom); $T_c=88$ K. This
shift of $T_j$ from 37 K to 40 K corresponds to the increase of
$I_c(T)$ by approximately 6.16 times [or equivalently, to the
decrease of $I_{a0}$ by 6.16 times at fixed $I_c(T)$] and leads to
the change of $T_d$ from $T_d\approx 38.5$ K to $T_d\approx 42$ K.
The solid lines show the numerical result obtained from
Eqs.~(\ref{14a})-(\ref{14e}), (\ref{51}), (\ref{52}) with
$U_0/T=30$, while the solid lines with dots are the analytical
result, Eq.~(\ref{58}). The horizontal dashed lines mark
$H_z^{(1)}$ in the Ohmic model when $I_c(T)=0$. The other dashed
lines show the analytical approximation obtained from
Eqs.~(\ref{12c}), (\ref{59b}). The temperature $T_x\approx  39$ K,
see Fig.~8.
 } \end{figure}   

In the nonlinear $R_b(I_b)$ model considered here the temperature
$T_d$ is of the order of $T_j$, and it does not coincide with the
temperature at which $R_b(T)\sim \mu_0\omega$, i.e., with the
appropriate temperature for the Ohmic model, $T_d^{\rm Ohmic}$
(for the parameters of Fig.~9, $R_b\sim \mu_0\omega$ at $T_d^{\rm
Ohmic}\approx 24$ K). The interesting feature of Fig.~9 is that
the dependences $H_z^{(1)}(T)$ obtained numerically from
Eqs.~(\ref{14a}) - (\ref{14e}), (\ref{51}), (\ref{52}) have
different behavior near $T_d$ for $T_j=37$ K and $T_j=40$ K. When
$T_j=40$ K, one has $\varphi_b\approx 0$ everywhere above
$T_d\approx T_j$, see Fig.~8. In other words, the inductance of
the edge wires can be disregarded completely. In this case
$H_z^{(1)}(T)$ has a nonmonotonic behavior near $T_d$. When
$T_j=37$ K, one has $T_d\approx 38.5$ K, and this $T_d$ lies in
the temperature region where the inductance of the edge wires is
essential. In this case the curves $H_z^{(1)}(T)$ of Fig.~9, in
fact, monotonically tend to zero in the vicinity of $T_d$. As the
boundary of the temperature region where the inductance of the
edge wires begins to play a role, one can take the point at which
Re$\tilde z_e = {\rm Im}\tilde z_e$ (i.e., at which
$\varphi_b\approx \pi/4$). This point is just the temperature
$T_x$ introduced for the Ohmic model.

In fact, expressions (\ref{58}), (\ref{59}) are applicable only
for $T>T_x$ when  $\varphi_b\approx 0$, and the dependence
$I_b^{(1)}(T)$ is determined by $I_c(T)$. In the opposite case, at
$T<T_x$ the phase shift $\varphi_b$ rapidly tends to $\pi/2$, and
expressions (\ref{58}), (\ref{59}) become small. This means that
the approximation used in deriving these expressions fails in this
region of temperatures. In other words, one cannot neglect the
time intervals when $I_b$ changes from $I_c$ to $-I_c$ or vice
versa. However, numerical analysis shows that near $T_d$, even
though $T_d<T_x$, the dependence $H_z^{(1)}(T)$ obtained from
Eqs.~(\ref{14a}) - (\ref{14e}), (\ref{51}), (\ref{52}) can be well
approximated by formula (\ref{12c}) if one takes
\begin{equation}\label{59b}  
I_b^{(1)}(T)\approx I_c(T),
\end{equation}
see Fig.~9. This statement remains true if instead of
Eq.~(\ref{59a}) one takes some other form for the temperature
dependence of $I_c$. It is only important that the function
$I_c(T)$ be sufficiently sharp near the point $T_j$. This finding
means that investigation of $H_z^{(1)}(T)$ near $T_d$ can give
information on the temperature dependence of the critical current
$I_c$ in this temperature region, see Appendix B.

\section{Conclusions}   

To describe the magnetic fields generated by the ac current in the
strip, we formally consider the strip as a set of the two edge
wires and of wires that form the bulk of the strip. The
resistances of all these wires can be nonlinear functions of the
currents flowing in them, and the resistances of the edge wires
$R_l$, $R_r$ differ from the resistances of the bulk ones. The
resistances of the edge wires characterize the edge barrier, and
they are generally different ($R_l\neq R_r$). To describe the
electrodynamics of this strip, we also take into account the
inductances of the wires. Equations (\ref{14a}) - (\ref{14f})
provide the description of the magnetic fields and currents in the
strip in this general case. We then consider in details three
special models which can be useful for understanding the
experimental data in various superconductors. The obtained results
can be summarized as follows:

{\it Linear (Ohmic) model}. In this case the resistances of all
the wires are assumed to be independent of the currents flowing in
them, and hence the electric fields in the strip are linear functions
of these currents. To describe the known experimental data, it is
necessary to assume that at $T_c$ the resistance of both edge
wires $R_e$ is larger than the bulk resistance $R_b$, but with
decreasing temperature $T$, $R_e$ decreases sharper than $R_b$.
The temperatures $T_{sb}$ and $T_d$ defined in Fig.~1 can then be
found from the conditions: $R_b(T_{sb})=R_e(T_{sb})$, and
$R_b(T_d)\approx \mu_0\omega$. The latter condition means that in
the Ohmic model at $T_d$ the ac magnetic field is expelled from
the strip due to the skin effect, and in the vicinity of this
$T_d$ the inductance of the bulk wires becomes essential. Since
$R_e(T)$ decreases sharper than $R_b(T)$ with decreasing $T$,  the
inductance of the edge wires begins to play a role at a
temperature $T_x$ that is higher than $T_d$. This $T_x$ is defined
by Re$\tilde z_e(T_x)\approx {\rm Im}\tilde z_e(T_x)$ where
$\tilde z_e$ is given by Eq.~(\ref{18b}).

We now recite the main features of the Ohmic model. The first
harmonic of the ac magnetic field is practically independent of
the asymmetry of the edge barrier, $\eta$, which is defined by the
asymmetry of the resistances $R_l$ and $R_r$, Eq.~(\ref{24a}),
while the second harmonic depends essentially on this asymmetry
and exists only at $\eta\neq 0$. The amplitude $H_z^{(1)}$ of the
first harmonic measured in phase with the applied ac current
depends on the frequency $\omega$ only near $T_d$, while the
magnitude of the second harmonic of the ac field changes with
$\omega$ only near $T_x$. At $T_x$ the magnitude of the second
harmonic sharply decreases, and a first harmonic that is
out-of-phase with the ac current appears. A characteristic
feature of the Ohmic model is also that the ac magnetic field
measured in units of $I_{a0}/w$, and hence the temperatures
$T_{sb}$, $T_x$, $T_d$, are independent of the amplitude $I_{a0}$
of the applied current.

{\it Model with nonlinear} $R_e$. In this model $R_e$ (edge) is
assumed to be a nonlinear function of the current $I_e$ flowing in
the edge wires. Namely, we assume the following: If $I_e$ exceeds
a critical current characterizing the edge barrier, the edge wires
do not differ from those in the bulk. But this critical current is
implied to increase with decreasing $T$, and when it exceeds
$I_e$, the resistance $R_e$ sharply drops. Thus, in contrast to
the Ohmic case, in this model a sharper decrease of $R_e(T)$ than
of $R_b(T)$ is explained by the nonlinear dependence of $R_e$ on
$I_e$. The temperature $T_{sb}$ is now near the temperature at
which the above-mentioned critical current reaches $I_e$. The
temperature dependence of the first harmonic $H_z^{(1)}$ is
qualitatively similar to that of the Ohmic model, and it does not
change essentially with the asymmetry of the edge barrier. On the
other hand, the second harmonic of the ac magnetic field is highly
sensitive to the asymmetry as in the Ohmic case, but now the
temperature where this harmonic sharply drops is due to the
temperature $T_1$ at which some characteristic current of the edge
barrier reaches $I_{a0}$. In this nonlinear model $T_1$ is
independent of the frequency $\omega$, and it plays the role of
the temperature $T_x$ marked in Fig.~1.

{\it Model with nonlinear} $R_b$. In this model $R_b$ (bulk) is
assumed to be a nonlinear function of the currents  flowing in the
bulk of the strip, namely: When the critical current for bulk
pinning, $I_c(T)$, which increases with decreasing $T$, exceeds
the bulk current $I_b$, the resistance $R_b$ sharply drops. Within
this model the temperature $T_d$ is not due to the skin effect as
it occurs for the Ohmic case, but now $T_d$ is close to the
temperature $T_j$ at which the critical current in the strip
$I_c(T)$ reaches the amplitude $I_{a0}$, $I_c(T_j)=I_{a0}$. Thus,
the characteristic feature of this model is that $T_d$ depends on
the applied amplitude $I_{a0}$ of the applied current rather than
on its frequency $\omega$. The temperature dependence of the first
harmonic $H_z^{(1)}$ near $T_d$ is mainly determined by the
dependence $I_c(T)$; this enables one to extract information on
the critical current in this temperature region from the measured
$H_z^{(1)}(T)$, see Fig.~9 and Appendix B. Interestingly, the
shape of $H_z^{(1)}(T)$ in the vicinity of $T_d$ is different for
the cases $T_d>T_x$ and $T_d<T_x$ where $T_x$ is defined by
Re$\tilde z_e(T_x)\approx {\rm Im}\tilde z_e(T_x)$, i.e., it marks
the point where the inductance of the edge wires becomes
important.

Different superconductors generally have different dependences
$R_b(T)$, $R_e(T)$, $I_c(T)$, and thus are characterized by
different values of the parameters that are essential in the
formulas of this paper. These parameters also depend on the
dimensions of the sample and frequency $\omega$. Therefore,
various situations may be expected to occur in experiments;
compare, e.g., the data of Refs.~\onlinecite{1} and
\onlinecite{3}. Possibly, some of them will not be well described
by the simple models considered above. However, the above results
allow one to understand the data for a specific experiment first
qualitatively and then to describe them quantitatively by an
improved model.

 \acknowledgments

  This work was supported by the German Israeli Research
Grant Agreement (GIF) No G-705-50.14/01.

\appendix

\section{Asymmetric nonlinear $R_e$ model}  

Here we present formulas for the first $H_z^{(1)}$ and the second
$H_z^{(2)}$ harmonics of the ac magnetic field in the framework of
the nonlinear model of Sec.~IV in the asymmetric case when
$I_{en}^c/ I_{ex}^c\ge R_{ex}/R_{en} \ge 1$. Let us introduce the
following notations: $q\equiv w/r$,
\begin{eqnarray}
I_{a1}\!\equiv \!I_{ex}^c\!\cdot\left (\!1+\!{R_{ex}\over R_{en}}+
\!{R_{ex} \over R_b}\right )\!,\ \ \ I_{a4}\!\equiv \!
I_{en}^c\!\cdot(2+q ), \nonumber
\end{eqnarray}
and
\begin{eqnarray}
I_{a2}\!\equiv \!I_{ex}^c\!+I_{en}^c\cdot\left (\!1+\!{R_{en}\over
R_b}\right )\!,\ \ \ I_{a3}\!\equiv \! I_{en}^c+I_{ex}^c\!\cdot
(\!1+q ), \nonumber
\end{eqnarray}
if $R_0/R_{en}>I_{en}^c/I_{ex}^c$, while if $R_0/R_{en}<
I_{en}^c/I_{ex}^c$, we define $I_{a2}$ and $I_{a3}$ as follows:
\begin{eqnarray}
I_{a2}\!\equiv \!I_{ex}^c\!\!\cdot\left (\!1\!+\!q\!+\!q{R_b\over
R_{en}}\!\right )\!,\ \ I_{a3}\!\equiv \!I_{en}^c\!\!\cdot\left
(\!1\!+\!{R_{en}\over R_b}\!+\!q{R_{en}\over R_b}\!\right )\!.
\nonumber
\end{eqnarray}
We also introduce $\phi_i\equiv \min({\rm
arcsin}[I_{ai}/I_{a0}],\pi/2)$ with $i=1-4$.

The first harmonic $H_z^{(1)}(x)$ is determined by $I_e^{(1)}$,
see Eq.~(\ref{12c}). This $I_e^{(1)}$ is described by the
following long but straightforward expression:
\begin{eqnarray}\label{A2}
 I_e^{(1)}\!\!\!\!&=&\!\!\!\frac{2I_{a0}}{\pi}\,{R_b\over
 R_b+R_e}\!\left (\!\phi_1\!-\!{\sin2\phi_1\over 2}\!\right)\!
 +\frac{4I_{ex}^c}{\pi}(\cos\phi_1 \nonumber \\
&-&\!\!\sigma_1\cos\phi_3\!-\sigma_2\cos\phi_2)\!
+\!\frac{4I_{en}^c}{\pi}(\sigma_1\cos\phi_2
+\!\sigma_2\cos\phi_3 \nonumber \\
&-&\cos\phi_4)\!+\frac{2I_{a0}}{\pi}\,{R_b\over R_b+R_{en}}\Big
(\phi_2-\phi_1-{\sin2\phi_2\over 2} \nonumber \\
&+&{\sin2\phi_1\over 2}\Big )-\frac{4I_{ex}^c}{\pi}\,{R_b\over
R_b+R_{en}}(\cos\phi_1- \cos\phi_2) \nonumber \\
&+&\!\frac{2I_{a0}}{\pi(1+q)}\!\left (\phi_4\!
-\!\phi_3\!+\!{\sin2\phi_3\over 2}\!-{\sin2\phi_4\over
2}\right)\!\! \nonumber \\
&-&\frac{4I_{en}^c}{\pi(1+q)}(\cos\phi_3 -\!\cos\phi_4)
+\frac{4I_{a0}}{\pi(2+q)}\Big ({\pi\over 2} -\!\phi_4\! \nonumber
\\
&+&\!{\sin2\phi_4\over 2}\!\Big )\! +\sigma_2\frac{2I_{a0}}{\pi}\,
{q R_b+ R_{en}\over q(R_b+R_{en})+R_{en}} \Big (\phi_3\!
-\!\phi_2 \nonumber \\
&+&\!{\sin2\phi_2\over 2}-{\sin2\phi_3\over 2}\Big )\,,
\end{eqnarray}
where $\sigma_1=1$ if $R_0/R_{en}>I_{en}^c/I_{ex}^c$ and
$\sigma_1=0$ otherwise, while $\sigma_2=1-\sigma_1$. In the
symmetric case when $I_{en}^c/ I_{ex}^c= R_{ex}/R_{en} = 1$, one
has $R_{en}=R_{ex}=2R_e$, $I_{en}^c=I_{ex}^c=I_e^c/2$,
$\sigma_2=0$, $I_{a1}=I_{a2}=I_1$, $I_{a3}=I_{a4}=I_2$ where the
currents $I_1$, $I_2$ are defined in Sec.~IV. In this case formula
(\ref{A2}) reduces to Eq.~(\ref{25}).

According to formula (\ref{12d}), the second harmonic is
determined by $\Delta I_e^{(2)}$. This $\Delta I_e^{(2)}$ is given
by:
\begin{eqnarray}\label{A4}
\Delta I_e^{(2)}\!\!\!\!&=&\!\!\!I_{a0}{(R_{ex}-R_{en})\over
(R_{ex}+R_{en})}\, {R_b\over R_b+R_e}\!\left (\!{1\over 3}
-\!{\cos\phi_1\over 2}+{\cos3\phi_1\over 6}\!\right)\!
\nonumber \\
&+&I_{a0}{R_b\over R_b+R_{en}}\Big ({\cos\phi_1\over 2}-
{\cos3\phi_1\over 6}-{\cos\phi_2\over 2} \nonumber \\
&+&{\cos3\phi_2\over 6}\Big )+I_{ex}^c{2R_b+R_{en}\over
R_b+R_{en}}\Big ({\sin2\phi_2\over 2}-{\sin2\phi_1\over 2}\Big
)\nonumber \\
&+&\!I_{en}^c\,{(2+q )\over 2(1+q )}(\sin2\phi_3 -\!\sin2\phi_4)
+\!{I_{a0}\over (1+q )}\Big ({\cos\phi_4\over 2} \nonumber \\
&-&\!\!\!{\cos3\phi_4\over 6}-\!{\cos\phi_3\over
2}+\!{\cos3\phi_3\over 6}\!\Big )\!+\!\sigma_1\!
{(\!I_{en}^c\!\!-\!I_{ex}^c)\over 2}(\sin2\phi_2 \nonumber \\
&-&\!\sin2\phi_3)+\sigma_2I_{a0}{q R_b-R_{en}\over
q(R_b+R_{en})+R_{en}}\Big ({\cos\phi_2\over 2} \nonumber \\
&-&{\cos3\phi_2\over 6}-{\cos\phi_3\over 2}+{\cos3\phi_3\over
6}\Big ).
\end{eqnarray}
In the symmetric case the quantity $\Delta I_e^{(2)}$ vanishes, as
it should be.

We now present formulas for the resistance $R$ of the strip in
this model. Such formulas may be useful to compare data obtained
from the ac experiments and from the conventional transport
measurements. Let us define four temperatures $T_i$ as
temperatures at which $\tilde I_{ai}(T_i)=1$, where $\tilde
I_{ai}(T)\equiv I_{ai}(T)/I_{a0}$, $i=1-4$, and $I_{a0}$ is the
applied transport current. Then, one has
\begin{eqnarray}\label{A5}
R\!&=&\!{R_bR_e\over R_b+R_e}\ \ \ \ \ \ \ \ \ \ \ \ \ \ \ \ \
{\rm for}\ T\le T_1,
\nonumber \\
R\!&=&\!{R_bR_{en}\over R_b+R_{en}}(1-\tilde I_{ex}^c)\ \ \ \ {\rm
for}\ T_1\le T\le T_2, \nonumber \\
R\!&=&\!\sigma_1R_b(1\!-\!\tilde I_{ex}^c\!-\!\tilde I_{en}^c)+
{\sigma_2q R_bR_{en}\over q(R_b\!+\!R_{en})\!+\!R_{en}} \nonumber \\
& &\ \ \ \ \ \ \ \ \ \ \ \ \ \ \ \ \ \ \ \ \ \ \ \ \ \ \ \ \
{\rm for}\ T_2\le T\le T_3, \\
R\!&=&\!R_b{q\over (1+q )}(1-\tilde
I_{en}^c)\ \ \ \ \ {\rm for}\ T_3\le T\le T_4, \nonumber \\
R\!&=&\!R_b{q\over (2+q )}\ \ \ \ \ \ \ \ \ \ \ \ \ \ \ \ \ {\rm
for}\ T_4\le T , \nonumber
\end{eqnarray}
where $\tilde I_{en}^c\equiv  I_{en}^c/I_{a0}$ and $\tilde
I_{ex}^c\equiv  I_{ex}^c/I_{a0}$. Note that $R$ is independent of
the applied current $I_{a0}$ at $T<T_1$ and at $T>T_4$. Besides
this, if $R_0/R_{en}=I_{en}^c/I_{ex}^c$ at some temperature $T_*$,
and if the interval $T_2 - T_3$ lies above $T_*$, a
current-independent resistance occurs in this interval as well.

\section{Extraction of $I_c(T)$ from experimental data}  

If the nonlinear $R_b$ model is applicable to describe some
experimental data on $H_z^{(1)}(T)$  near the temperature $T_d$,
the critical current $I_c(T_d)$ can be approximately estimated
from the relations $I_c(T_j)=I_{a0}$ and $T_d\approx T_j$. Changing
$I_{a0}$, one finds $I_c(T)$ in some temperature interval which is
determined by the range of $I_{a0}$. On the other hand, the
temperature dependence $I_c(T)$ can be also estimated from the
shape of the curves $H_z^{(1)}(T)$ measured at fixed $I_{a0}$.
Using the approximation (\ref{59b}) and formula (\ref{12c}), we
obtain
\begin{equation}\label{B1}
 I_c(T)={2w H_z^{(1)}(T)- I_{a0}\cdot g \over f-g},
\end{equation}
where $H_z^{(1)}(T)$ is the first harmonic measured at an inner
point $x$ (i.e., at $|x|< w$), $g=(xw/\pi)/(w^2-x^2)$, $f=2w
H_z^{(1)}(T_c)/ I_{a0}$, and $H_z^{(1)}(T_c)$ is the first
harmonic $H_z^{(1)}$ measured at the same point $x$ at a
temperature near $T_c$. Here, to express $f$ in terms of
$H_z^{(1)}(T_c)$, we have used formula (\ref{12c}) again and have
taken into account that at $T=T_c$ the edge barrier is absent, and
$I_b^{(1)}=I_{a0}$. When the sample is an ``ideal'' strip (i.e.,
uniform with constant thickness, width, and homogeneous critical
current density), one has $f=(1/2\pi)\ln(|w-x|/|w+x|)$. Note that
these two ways of estimating $I_c$ more reliably give the
temperature dependence of $I_c$ rather than the {\it absolute}
value of the critical current. Indeed, at $T=T_d$ when one has
$H_z^{(1)}(T)=0$, the first method gives $I_c=I_{a0}$, while the
second method leads to $I_c=I_{a0}\cdot g/(g-f)\approx I_{a0}/2$.
This discrepancy is caused by the inaccuracy of the relation
$T_d\approx T_j$ and of the approximation (\ref{59b}).

{}


\begin{thebibliography}{}
\bibitem{c1} However, we shall still
assume that there is no skin effect across the thickness of the
strip $d$. This is not a restrictive assumption since the ac
magnetic field $H_z$ is completely expelled from the sample before
such effect occurs at lower temperatures.
\bibitem{c} Beside this, if one reaches $U_e(0)-U_b(0)\sim 4$ by
changing $R_e^0$, the shapes of the curves Re$H_z^{(1)}(T)$ differ
noticeably from those of Fig.~1.
\bibitem{c3} Note that Fig.~7 has been plotted for the realistic
current-voltage law, Eqs.~(\ref{51}) and (\ref{52}), rather than
for the idealized model (\ref{53}).
\end{thebibliography}
\end{document}